\documentstyle[prl,aps,epsf,preprint]{revtex}

\begin{document}

\draft


\title{Hall-effect in LuNi$_2$B$_2$C  
in normal and superconducting mixed states}

\author{V.N. Narozhnyi$^{a,b,c,}$\cite{PresAdd}, 
J. Freudenberger$^b$, V.N. Kochetkov$^{a,c}$, K.A. Nenkov$^{b,c}$, 
G. Fuchs$^b$,  K.-H. M\"uller$^b$}

\address{$^a$Institute for High Pressure Physics, Russian Academy of 
Sciences, Troitsk, Moscow Region, 142092, Russia} 
\address{$^b$Institut f\"ur Festk\"orper- und Werkstofforschung 
Dresden e.V., Postfach 270016, D-01171 Dresden, Germany}
\address{$^c$International Lab. of High Magnetic Fields and Low 
Temperatures, Gajowicka 95, 53-529 Wroclaw, Poland} 

\date{\today}

\maketitle

\begin{abstract}
The Hall resistivity $\rho_{xy}$ of $\rm LuNi_2B_2C$ is negative in 
the normal as well as in the mixed state and has no sign reversal 
typical for high-$T_c$ superconductors. A distinct nonlinearity in 
the $\rho_{xy}$ dependence on field $H$ was found in the normal state 
for $T\lesssim$ 40~K, accompanied by a large magnetoresistance 
reaching +90\% for $\mu_0H=16$~T at $T=20$~K. The scaling relation 
$\rho_{xy}\sim\rho_{xx}^\beta$ ($\rho_{xx}$ is the longitudinal 
resistivity) was found in the mixed state, the value of $\beta$ being 
dependent on the degree of disorder.

\end{abstract} 

\vskip1pc

\pacs{$\em Keywords$: A. superconductors, D. electronic transport,
D. galvanomagnetic effects.}


\narrowtext

Investigations of the Hall effect in the normal and superconducting 
(SC) mixed states give an important information about the electronic 
structure and the vortex dynamics. 
The nature of both of them is not settled yet for the recently 
discovered \cite{nagaraj,cava} SC borocarbides $R\rm Ni_2B_2C$ 
($R$=Y, rare earth). Despite the fact that the borocarbides have a 
tetragonal layered crystal structure, their electronic properties 
indicate three-dimensionality showing little anisotropy 
\cite{lee,pickett,kim}. Interesting features of some borocarbides 
($R$=Lu, Er) are their unusual square vortex lattice \cite{dewilde} 
and, connected with that, the anisotropy of the critical magnetic 
field $H_{c_2}(T)$ in the $ab$-plane \cite{metlushko,rath}. Since the 
Hall effect in the SC state may depend on peculiarities of the vortex 
lattice, its study is of special interest for systems with an 
anomalous square vortex lattice. To our knowledge, no data on the 
Hall effect for $\rm LuNi_2B_2C$ and only few for other borocarbides 
are known so far \cite{fischer,narozh_jltp,mandal,narozh_jac}. For 
the compounds based on $R$ = Y \cite{fischer,narozh_jltp,mandal}, Ho 
\cite{fischer,mandal}, La \cite{fischer}, and Gd \cite{mandal}, it 
was observed that the normal state Hall coefficients $R_H$ are 
negative and only slightly temperature dependent. At the same time, a 
negative but strongly temperature-dependent $R_H$ was found for the
heavy-fermion-like compound $\rm YbNi_2B_2C$ \cite{narozh_jac}. No 
sign reversal of the Hall resistivity $\rho_{xy}$ typical for 
high-$T_c$ superconductors was observed for $\rm YNi_2B_2C$ 
\cite{narozh_jltp}, but the mixed state Hall effect was not studied 
for borocarbides systematically.

The mixed state Hall effect was investigated mainly for high-$T_c$ 
SC. Two unexpected effects were found: sign reversal of $\rho_{xy}$ 
for $T<T_c$ and a striking scaling relationship between $\rho_{xy}$ 
and the longitudinal resistivity $\rho_{xx}$ in the SC transition 
region, $\rho_{xy}\sim \rho_{xx}^\beta$.  $\it{Sign~reversal}$ of 
$\rho_{xy}$ has been observed for several types of high-$T_c$ SC, 
e.g., $\rm YBa_2Cu_3O_{7-y}$ \cite{galffy,ri}, $\rm 
Bi_2Sr_2CaCu_2O_8$ \cite{ri,samoilov_93}, $\rm Tl_2Ba_2CaCu_2O_8$ 
\cite{budhani} as well as for some low-$T_c$ SC:  In-Pb alloys, V, Nb 
(see \cite{parks}).  Several models have been proposed for the 
description of this effect \cite{wang_prl91,dorsey_pr92,kopnin}, but 
its nature is not yet fully understood.  Meanwhile, in 
\cite{dorsey_pr92,kopnin}, the sign reversal of $\rho_{xy}$ below 
$T_c$ is expected to be not a universal property, but crucially 
dependent on the shape of the Fermi surface.  
$\it{Scaling~behavior}$, $\rho_{xy}$$\sim$$\rho_{xx}^\beta$, in the 
SC state was observed e.g. for $\rm YBa_2Cu_3O_{7-y}$ ($\beta$=1.7) 
\cite{luo}, $\rm Bi_2Sr_2CaCu_2O_8$ ($\beta$$\approx$2) 
\cite{samoilov_93}, and $\rm Tl_2Ba_2CaCu_2O_8$ ($\beta$$\approx$2) 
\cite{budhani}. This has been interpreted \cite{dorsey_fischer} in 
the framework of glassy scaling near a vortex-glass transition.  
Considering the effect of flux pinning on the Hall conductivity, 
$\sigma_{xy}$, it is expected that $\sigma_{xy}$ is independent of
the degree of disorder \cite{vinokur_prl93}. Scaling behavior, 
$\rho_{xy}=A\rho_{xx}^\beta$ with $\beta=2$, is believed to be a 
general feature of any vortex state with disorder-dominated dynamics 
\cite{vinokur_prl93}. On the other hand Wang, Dong and Ting (WDT) 
\cite{wang_prl94} developed a theory for the Hall effect including 
both pinning and thermal fluctuations. Scaling and sign reversal of 
$\rho_{xy}$ are explained by taking into account the backflow current 
due to pinning \cite{wang_prl94}. Thereby, $\beta$ changes from 2 to 
1.5 as the pinning strength increases \cite{wang_prl94}.  
Controversial experimental results have been reported on the 
influence of disorder  on the mixed state Hall effect. Thus, for $\rm 
Tl_2Ba_2CaCu_2O_8$ irradiated by heavy ions, $\beta=1.85$ holds even 
after irradiation \cite{budhani}. On the other hand, for irradiated 
$\rm YBa_2Cu_3O_{7-y}$ samples, $\beta$ was found to be $1.5\pm 0.1$ 
compared to $2\pm 0.2$ for unirradiated ones \cite{kang} in 
accordance with WDT. 

In this communication, we report the Hall effect study for $\rm 
LuNi_2B_2C$. In the normal state, a negative and weakly temperature 
dependent $R_H$ was observed.  Contrary to the case of $\rm 
YNi_2B_2C$, the $\rho_{xy}(H)$ dependence was found to be essentially 
nonlinear for $T\lesssim 40$~K, accompanied by a very large 
magnetoresistance MR. In the SC transition region, scaling behavior 
was found, but no sign reversal of $\rho_{xy}$ was observed. The 
scaling exponent $\beta$ was found to be dependent on the degree of 
disorder and can be varied by annealing. This is attributed to a 
variation of the strength of flux pinning.

Polycrystalline $\rm LuNi_2B_2C$ (in the following denoted as PC AN) 
was prepared by arc-melting in Ar atmosphere and subsequent annealing 
at 1100$\rm\ ^\circ C$, as described in more detail in 
\cite{muller_96}. The phase purity of the samples was checked by 
X-ray diffraction. The reflexes reveal practically a single phase. 
For comparison, some measurements were performed on an unannealed 
sample (PC UNAN). Hall contacts with typical misalignment of less 
than 0.1 mm were used (typical dimentions of the samples were $ \rm 
3\times 1\times 0.3~mm^3$). The Hall voltage was measured for two 
directions of the field $H$. MR was measured by standard four-probe 
method. Most measurements were done at $\mu_0H$ up to 5~T, but some of 
them up to 16~T in different installation.

\begin{figure} 
\epsfysize=5.3cm 
\centerline{\epsfbox{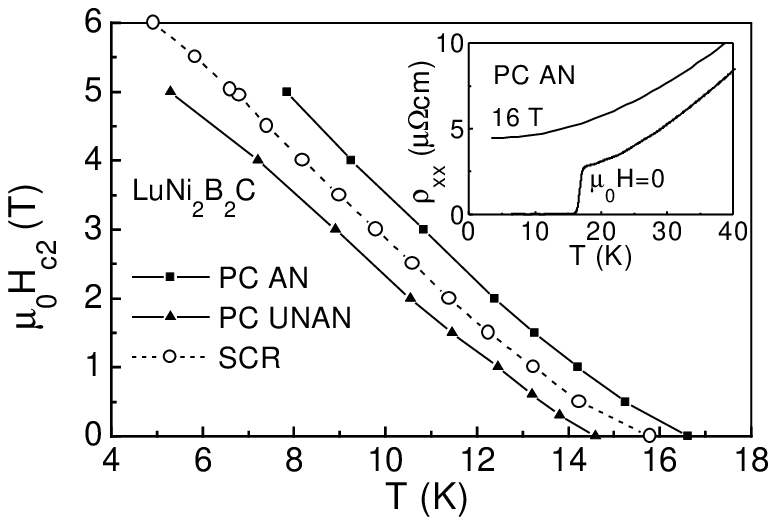}} 
\vspace{0.5pc}
\caption{$H_{c_2}$ vs. $T$ for three $\rm LuNi_2B_2C$ samples. Open 
symbols - the results in Ref. \protect \cite{metlushko} for single 
crystal.  Lines are guides for eye. The inset shows $\rho_{xx}$ vs. 
$T$ for the annealed polycrystalline sample.} 
\label{fig1} 
\end{figure}

The $\rho_{xx}$ vs. $T$ curves for PC AN are shown on the 
inset of Fig.\ \ref{fig1}.  The value of $\rho_{xx}$(17~K) is 
2.7~$\mu\Omega$cm, which is comparable with that of $\rm LuNi_2B_2C$ 
single crystals (1.9~$\mu\Omega$cm \cite{rath} and 2.5~$\mu\Omega$cm 
\cite{shulga}). The residual resistance ratio 
RRR=$\rho_{xx}$(300~K)/$\rho_{xx}$(17~K) is 41 for PC AN which is 
significantly higher than for single crystals (25 \cite{rath} and 
27 \cite{shulga}). The value of $T_c$=16.7~K is also slightly 
higher than that reported for single crystals (15.8~K 
\cite{metlushko}, 16.1~K \cite{rath}, 16.5~K \cite{shulga}). The 
width of the SC transition is 0.27~K which is close to the values 
0.2$\div$0.25~K typical for single crystals 
\cite{metlushko,rath,shulga}. These results give evidence for a good 
quality of the PC AN sample. PC UNAN has the lower value of $T_c$ 
(14.7~K) and an order of magnitude higher value of $\rho_{xx}(17~K)$ 
(not shown in a figure). 

The results for $H_{c_2}(T)$ are depicted on Fig.\ \ref{fig1}.  
($H_{c_2}$ was determined as in \cite{metlushko} by the 
extrapolation of the $\em{ac}$-susceptibility curve to zero 
susceptibility value.) For comparison, the data from \cite{metlushko} 
for a single crystal (SCR) with $H\parallel$$<$110$>$ are also shown 
(as open circles). The upper curvature (UC) in $H_{c_2}(T)$, near 
$T_c$, is clearly visible.  Note that, in accordance with 
\cite{shulga}, the UC region is smaller for the unannealed sample.  

\begin{figure} 
\epsfysize=10.8cm 
\centerline{\epsfbox{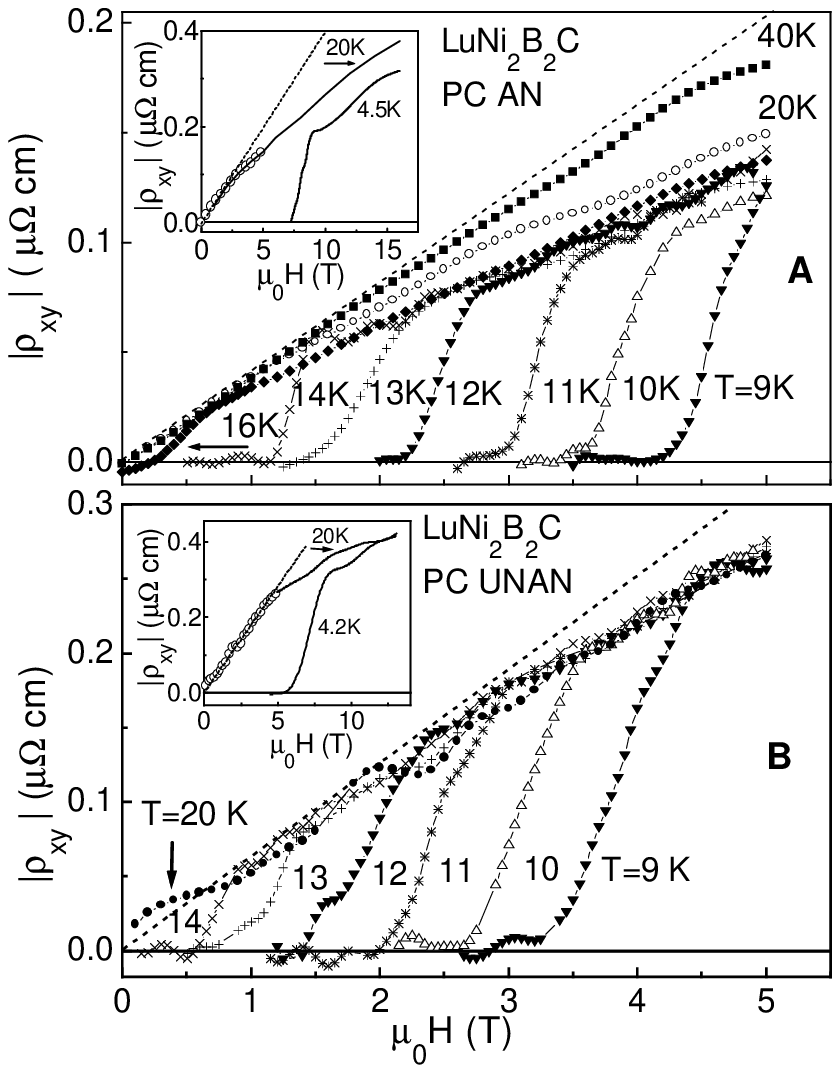}} 
\vspace{0.5pc}
\caption{$|\rho_{xy}|$ vs. $H$ for the annealed (A) and
unannealed (B) samples. The dashed lines are low-field asymptotes 
to the normal state curves. Open points in the insets denote the 
results obtained for $\mu_0H\leq$ 5~T.} 
\label{fig2} 
\end{figure}

The Hall resistivity $\rho_{xy}(H)$ of both samples in normal 
and SC states is shown in Fig.\ \ref{fig2}. At 
3.3~K$\leq T\leq$300~K, $\rho_{xy}$ is negative and has no sign 
reversal. 

$\it{In~the~normal~state}$, a pronounced nonlinearity in the
$\rho_{xy}(H)$ dependences is evident for $T\lesssim$ 40~K,  
which is more clearly seen in the insets of Fig.\ \ref{fig2} 
where some results obtained for $H$ up to 16~T are presented.
It should be underlined that no nonlinearlity in 
$\rho_{xy}(H)$ was reported for $\rm YNi_2B_2C$ \cite{narozh_jltp} 
and $\rm YbNi_2B_2C$ \cite{narozh_jac}. 

\begin{figure}
\epsfysize=3.1cm 
\centerline{\epsfbox{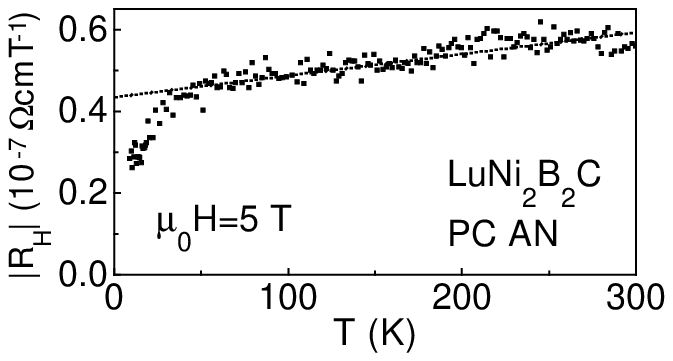}} 
\vspace{0.5pc}
\caption{Absolute value of $R_H$ vs. $T$ for $\rm LuNi_2B_2C$.} 
\label{fig3} 
\end{figure}

In Fig.\ \ref{fig3}, the $R_H(T)$ dependence of the PC AN sample 
is shown for $\mu_0H$=5~T. At $T \lesssim 40$~K, $R_H(T)$ deviates
from the linear behavior observed for high $T$ (dotted line).
This deviation is connected with the nonlinearity in 
$\rho_{xy}(H)$ at low $T$, shown in Fig.\ \ref{fig2}.  The 
value of $R_H$ is comparable with those reported for $\rm YNi_2B_2C$ 
\cite{fischer,narozh_jltp,mandal}, but it is five times smaller than 
the value $R_H(T)$=3$\cdot 10^{-7}\Omega$cm/T obtained
from band structure calculations \cite{pickett}. These deviations may 
be caused by correlation effects in borocarbides. The evaluation of 
the carrier density from the $R_H$ value at $T$=300~K, by using a 
single band model which is a rough approximation, gives 1.5 carriers 
per unit cell.  
 
\begin{figure} 
\epsfysize=5.7cm 
\centerline{\epsfbox{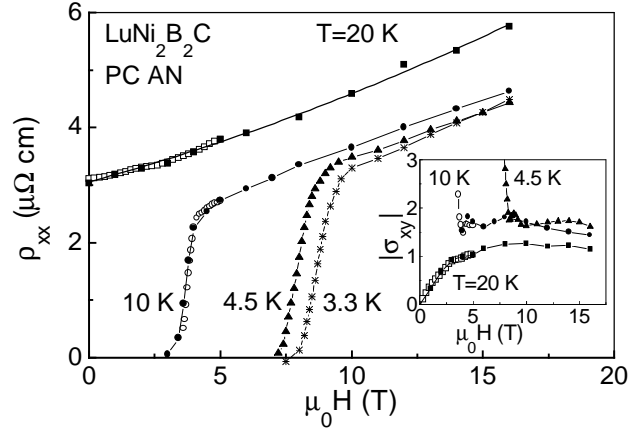}} 
\vspace{0.5pc}
\caption{Magnetoresistance for annealed $\rm LuNi_2B_2C$.  In the 
inset $\sigma_{xy}$ (in $10^{-2}\mu\Omega^{-1}cm^{-1}$) vs. 
magnetic field is shown. Open points denote the results obtained for 
$\mu_0H\leq$ 5~T. Lines are guides for eye.} 
\label{fig4} 
\end{figure}

As can be seen from Fig.\ \ref{fig4}, the values of 
MR=$(\rho_{xx}(H)-\rho_{xx}(0))/\rho_{xx}(0)$ for the PC AN sample at 
$T=20$~K are as high as 25 and 90\% for $\mu_0H=5$ and 16~T, 
respectively \cite{freud}. Note that an MR of only $\approx$7.3\% was 
observed, at $\mu_0H$=4.5~T and $T=20$~K, for a $\rm LuNi_2B_2C$ single 
crystal with RRR=25 ($H$ parallel to the tetragonal $c$-axis) 
\cite{rath}. A possible reason for the significantly larger MR for 
the polycrystalline sample compared to the single crystal \cite{rath} 
is the formation of open orbits on the Fermi surface of $\rm 
LuNi_2B_2C$ for $H$$\perp$$c$. (The possibility of the open orbits 
formation for borocarbides was pointed out in band structure 
calculations \cite{kim,lee}. It was claimed that one part of the 
Fermi surface is a cylinder along the $c$-axis \cite{kim}, so open 
orbits can be expected for $H$$\perp$$c$.) It is well known 
\cite{lifshitz} that open orbits can lead to large values of MR~$\sim 
H^2$, whereas closed orbits should give rise to saturation of MR($H$) 
for large $H$. In that case the averaging of MR should lead to a 
practically linear $\rho(H)$ dependence for polycrystals 
\cite{lifshitz} that should be {\em stronger} than that observed for 
single crystals for $H$$\parallel$$c$ when only closed orbits could 
be expected. Therefore, significantly larger MR observed for the 
polycrystals compared to one for the single crystal for 
$H$$\parallel$$c$ can be considered as an indication for the open 
orbits formation in $\rm LuNi_2B_2C$ for $H$$\perp$$c$. Measurements 
of the MR in high fields for single crystals with two configurations 
((i) $I\parallel c$ and $H\perp c$ and (ii) $I\perp c$ and 
$H\parallel c$) are necessary to check this hypothesis.

The nonlinear $\rho_{xy}(H)$ dependence and the large MR, found by 
us, as well as the anisotropy of $H_{c_2}$ in the $ab$-plane and the 
square vortex lattice, reported for $\rm LuNi_2B_2C$ at high $H$ 
earlier \cite{dewilde}, may be caused by the peculiarities of its 
electronic structure, because for $\rm YNi_2B_2C$ all these anomalies 
are absent. (For $\rm YNi_2B_2C$ a linear $\rho_{xy}(H)$ dependence, 
a substantially smaller MR \cite{narozh_jltp,freud}, and only a very 
small anisotropy of $H_{c_2}(T)$ \cite{rath} were observed.) These 
distinctions are probably connected with the difference between the 
Fermi surfaces of the two compounds.  For the borocarbides, the Fermi 
surface topology is very sensitive to the position of the Fermi level 
\cite{lee}, that may be slightly different for the two cases, Lu and 
Y, due to, e.g., different lattice constants. From our results 
\cite{freud} it follows, that the formation of open orbits is
probably easier in case of $\rm LuNi_2B_2C$.

As shown in the inset of Fig.\ \ref{fig4}, the nonlinearity in 
$\sigma_{xy}(H)$ is even more pronounced than for 
$\rho_{xy}(H)$ ($\sigma_{xy}\cong\rho_{xy}/\rho_{xx}^2$,
$\rho_{xx}>>|\rho_{xy}|$). In particular, $\sigma_{xy}$ is 
practically independent of $H$ for $\mu_0H=8\div$16~T, 
at $T=4.5\div$20~K. The nonlinear $\rho_{xy}(H)$ 
dependence and the large MR of $\rm LuNi_2B_2C$ are probably closely 
connected and result in a practically constant $\sigma_{xy}(H)$ for 
high fields. The reason why  $\sigma_{xy}$ is independent of $H$ 
for high fields, resulting in $\rho_{xy}\sim\rho_{xx}^2$ 
in the $\it{normal~state}$ is not yet understood. (Noteworthy, 
$\rho_{xy}\sim \rho_{xx}^2$ scaling in the $\it{normal}$ state
was observed also for the SC heavy fermion compound $\rm UBe_{13}$ 
\cite{aleks}.) A nonlinear $\rho_{xy}(H)$ dependence can also be 
obtained within the two-band model used, e.g., to interpret the 
nonlinear and even nonmonotonous $\rho_{xy}(H)$ dependence for 
$\rm UB_{13}$  \cite{aleks}. In this model, at low $H$, the light 
carriers with high mobilities give the prevalent contribution to 
$\rho_{xy}(H)$, whereas in high fields the contribution of the 
heavier carriers are more significant. Recently, in a different 
multi-band model \cite{shulga}, proposed for the 
$\rm Lu$- and $\rm Y$-borocarbides, the existence of at least two 
bands with significantly different Fermi-velocities was found 
to be essential for the quantitative description of $H_{c_2}(T)$ 
curves. Note that several groups of carriers with different masses 
have been directly observed for $\rm YNi_2B_2C$ in dHvA experiments 
\cite{nguyen}.

\begin{figure} 
\epsfysize=13.9cm 
\centerline{\epsfbox{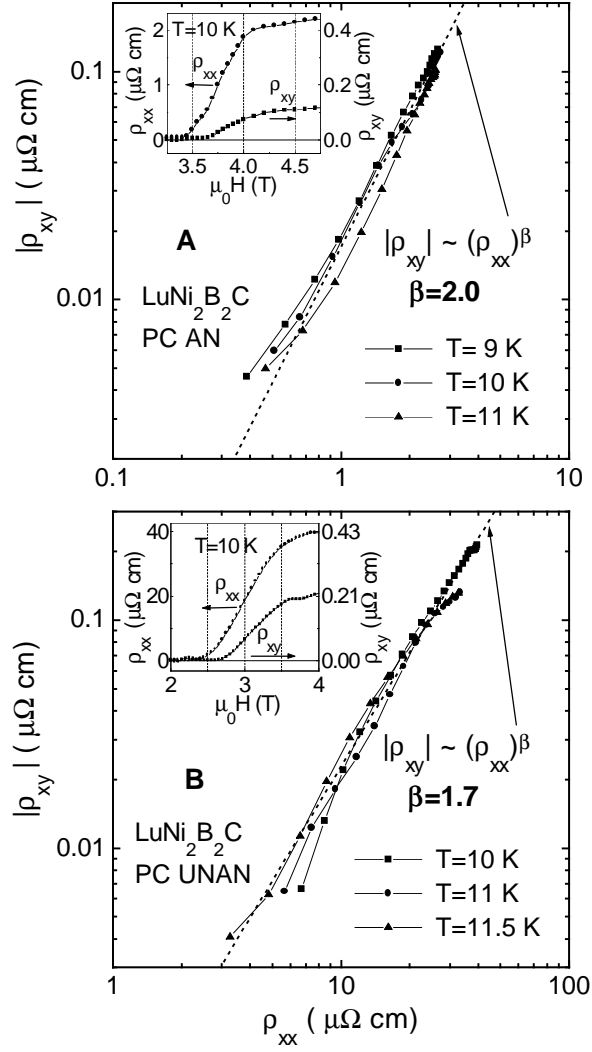}} 
\vspace{0.5pc}
\caption{$\mid\rho_{xy}\mid$ vs. $\rho_{xx}$ for the annealed (A) and 
unannealed (B) samples. In the insets $\mid\rho_{xy}\mid$ and 
$\rho_{xx}$ vs. magnetic field are simultaneously shown.} 
\label{fig5} 
\end{figure}

$\it{In~the~mixed~state}$, two regions concerning the behavior of 
$\rho_{xy}$ and $\rho_{xx}$  can be distinguished 
(see Fig.\ \ref{fig5}). For higher values of $H$ (or $|\rho_{xy}|$ 
or $\rho_{xx}$) it is clearly seen that the scaling behavior 
$|\rho_{xy}|=A\rho_{xx}^\beta$ holds for both samples. 
The value of $\beta$ is $2\pm$0.1 for PC AN and it 
decreases to 1.7$\pm$0.1 for PC UNAN that has an order of 
magnitude higher resistivity. This may be connected, in accordance 
with WDT, with an increase of pinning effects. For decreasing fields,
$\rho_{xx}$ vanishes at lower values of $H$ than $|\rho_{xy}|$ (see
the insets of Fig.\ \ref{fig5}), which is also consistent with 
theoretical predictions in \cite{wang_prl94}. Obviously, pinning 
effects are essential for the Hall effect in $\rm LuNi_2B_2C$.

In order to understand the absence of sign reversal in $\rho_{xy}$ 
for $\rm LuNi_2B_2C$, the following physical picture of the Hall 
effect in the mixed state \cite{dorsey_pr92,kopnin} can be used in 
addition to WDT theory:  there are two contributions to $\sigma_{xy}$ 
in the SC state, $\sigma_{xy}=\sigma_n+\sigma_{sc}$, where $\sigma_n$ 
is connected with normal carriers that experience a Lorentz force in 
the vortex core and $\sigma_{sc}$ is an anomalous contribution 
connected with the motion of vortices parallel to the electrical 
current $I$. In \cite{dorsey_pr92} and \cite{kopnin} it was claimed 
that $\sigma_{sc}\sim1/H$ and could have a sign opposite to that of 
$\sigma_n$. Therefore, in small $H$, the $\sigma_{sc}(H)$ term is 
more essential but in higher fields $\sigma_n(H)$ becomes to be 
dominant. If $\sigma_{sc}$ has a different sign than $\sigma_n$ it is 
possible to observe sign reversal in $\rho_{xy}(H)$ at $T<T_c$ 
\cite{dorsey_pr92,kopnin}. The relation 
$\sigma_{xy}=\sigma_n+\sigma_{sc}$ was verified and the term 
$\sigma_{sc}\sim1/H$ was observed in \cite{ginsberg} for YBCO. For 
$\rm LuNi_2B_2C$ the conductivity decreases with
increasing $H$ as can be seen, for $T$=10 and 4.5~K, from the inset 
of Fig.\ \ref{fig4}. It should be pointed out, that the observed 
$\sigma_{xy}$ vs. $H$ curves seem to change more rapidly, than $1/H$. 
A similar behaviour was observed for cuprates, see, e.g., Ref.\ 
\cite{matsuda_prb95}. Therefore the mechanism of the mixed-state 
Hall effect connected with vortex motion is assumed to work for 
borocarbides as well. The signs of $\sigma_n$ and $\sigma_{sc}$ are 
the same, contrary to the case of high-$T_c$ superconductors.  This 
may be the reason for the absence of sign reversal of $\sigma_{xy}$ 
in borocarbides.

In conclusion, we have investigated the Hall 
effect for $\rm LuNi_2B_2C$ in normal and mixed states. A negative 
and only slightly temperature-dependent $R_H$ was found for $T>T_c$.
The value of $R_H$ is five times smaller than that resulting from 
band structure calculations \cite{pickett}. Pronounced nonlinearity 
in the $\rho_{xy}(H)$ dependence was found in the normal state for 
$T\lesssim$ 40~K accompanied by a very large MR.  The possibility of 
open orbits formation on the Fermi surface for $H\perp c$ is pointed 
out. In the mixed state scaling behavior, 
$\rho_{xy}\sim\rho_{xx}^\beta$, was observed but no sign reversal 
typical for high-$T_c$ superconductors was found. The scaling 
exponent $\beta$ is $2\pm 0.1$ for the annealed sample with low 
$\rho_{xx}$ and it decreases to $1.7\pm 0.1$ for the unannealed one. 
This is attributed to a variation of the strength of flux pinning. 

We thank S.-L. Drechsler, V.V. Marchenkov and V.I. Nizhankovskii 
for discussions. This work was supported by RFBR grant 96-02-00046G, 
DFG grant MU1015/4-1.

\begin{references}
\bibitem{freud}Results on the MR of $\rm LuNi_2B_2C$ and $\rm 
YNi_2B_2C$ will be discussed in more detail elsewhere.
\end{references}
\end{document}